\documentclass[11pt,twoside]{article}
\usepackage{newpasp}

\def\edcomment#1{\iffalse\marginpar{\raggedright\sl#1\/}\else\relax\fi}
\reversemarginpar

\begin{document}
\title{Chandra X-ray Observations of Cygnus A and Pictor A}
\author{A. S. Wilson}
\affil{Astronomy Department, University of Maryland, College Park, MD 20742,
U.S.A. and Space Telescope Science Institute, 3700 San Martin Drive, Baltimore,
MD 21218, U.S.A.}
\author{A. J. Young}
\affil{Astronomy Department, University of Maryland, College Park, MD 20742,
U.S.A.}
\author{P. L. Shopbell}
\affil{Department of Astronomy, California Institute of Technology, MS 105-24,
Pasadena, CA 91125}

\begin{abstract}
Results from Chandra observations of the two nearest, powerful radio
galaxies are summarised.
\end{abstract}

\section{Introduction}
The talk described observations of Cygnus A (z = 0.0562) and Pictor A
(z = 0.035) with the Chandra X-ray Observatory. Since the early
results from these observations are published (Wilson, Young \&
Shopbell 2000, 2001), we here confine ourselves to a summary of the
main conclusions and a few remarks about other aspects of the results
on Cygnus A, which will be amplified in a future paper.

\section{Cygnus A}
\subsection{Cavity, Hot Spots, Jets and Nucleus}

The Chandra image of Cygnus A is shown in Fig. 1. This figure shows
only the region of the radio source and omits the large-scale, X-ray
emitting gas which comprises the intracluster medium. The dominant
feature of Fig. 1 is a limb-brightened, elliptical (presumably
prolate-spheroidal in three dimensions) structure. The major axis
coincides with that of the radio source and the radio hot spots are
seen as compact X-ray sources at the ends of the major axis. This
structure appears to be the observational manifestation of the cavity
which is expected to be inflated in the intracluster medium by
relativistic material which has passed through the hot spots at the
ends of the jets. Its existence and dynamics were first inferred and
explored in the classic paper by Scheuer (1974). The ratio of major to
minor axis of the ellipse is $\sim$ 2 -- 2.5, in agreement with the
expectations of simple models (cf.  Begelman \& Cioffi 1989). Bright,
curved bands of emission project along the minor axis of the
ellipse. It is tempting to interpret these features as ``belts'' of
gas extending around the equator of the prolate spheroidal structure.

A curved structure runs along the major axis from the nucleus to each
pair of hot spots. These two features are apparently the X-ray
manifestations of the jets. They are, however, broad, diffuse and of
low contrast, quite unlike the sharply defined radio structure of the
large scale NW (the near side of the radio source) jet (cf. Carilli et
al. 1996). The X-ray jet almost disappears near the NW hot spots (A
and B in the nomenclature of Hargrave \& Ryle 1974).  There is also an
faint elliptical semi-ring in the X-ray image; this semi-ring extends
$\simeq$ 10$^{\prime\prime}$ to the NE of hot spots A and B. The
combination of this structure and the two hot spots resembles a set of
`headphones' (the hot spots mark the `earpieces'). Definitive
statements cannot be made in view of the low contrast of the X-ray
jets, but we speculate that either:

\noindent
a) the NW jet is continuous and steady but much broader than would be
inferred from the radio maps.  In this interpretation, the hot spots
and the semi-ring reflect the interaction of the outer annulus of the
jet with the intracluster gas. The observed radio jets and hot spots
might then be effects of limb-brightening at the edges of the broad
jet (cf. Blandford [1996] who suggests that the hot spots might be a
ring vortex seen in projection).  However, the great brightness of the
hot spots and the close agreement between their radio synchrotron and
X-ray synchrotron self-Compton emissions (see next subsection)
challenge a limb-brightening interpretation.  Alternatively, an
absence of radio emission from the center of the jet may mean that the
jet is actually hollow or that the jet's center comprises material
unconducive to radio emission.

\noindent
or: b) the NW jet is intrinsically narrow, as inferred from the radio
maps, but the direction in which it is launched is continuously
changing (precession of the nozzle?), so that its terminus traces out
a circle where it meets the intracluster gas. The elliptical semi-ring
is then the projection of this terminal circle, and the hot spots are
two locations along the circle where the radio emission is unusually
bright, as a result of projection or other effects.

The X-ray nucleus is extended in the SE - NW direction. Its spectrum
is shown in Fig. 2. The high energy continuum may be modeled as an
absorbed power law, as known from previous work (e.g. Arnaud
1996). Notable features include: a) a K$\alpha$ line from neutral or
lowly ionized iron (seen in the observed spectrum near 6.0 keV), b)
strong K edge absorption from neutral or lowly ionized iron (observed
near 7.0 keV), and c) emission at energies below 2 keV in excess of
that expected from the absorbed power law. Emission lines, which are
confirmed in an independent spectrum with higher signal to noise at
low energies, are apparent and tentatively identified with K$\alpha$
transitions of silicon, neon and possibly magnesium. The soft X-ray
emission apparently originates from an extended region viewed through
a relatively low absorbing column.

\subsection{Synchrotron Self-Compton Emission from the Hot Spots}

Fig. 3 (left) shows X-ray contours on a grey scale of a 6 cm image
with resolution 0\farcs35 (Perley, Dreher \& Cowan 1984) in the
vicinities of the western hot spots (A and B). The extents and
morphologies of the X-ray and radio hot spots are very similar, with
the directions of elongation agreeing to within a few degrees. Similar
results are obtained for hot spots D and E.  The X-ray spectra of hot
spots A and D have been modeled with an absorbed power law. In both
cases, the absorbing column is N$_{\rm H}$ = 3.3 $\times$ 10$^{21}$
cm$^{-2}$, in excellent agreement with the Galactic column in the
direction of Cygnus A. The photon indices are also similar at $\Gamma$
= 1.8 $\pm$ 0.2.

Thermal models for the X-ray emission may be ruled out as they require
too high gas densities. Fig. 3 (right) shows the results of
calculation of a synchrotron self-Compton (SSC) model for hot spot
A. As may be seen, the predicted SSC radiation is in excellent
agreement with the Chandra-observed spectrum for a magnetic field of
1.5 $\times$ 10$^{-4}$ gauss. This value may be compared with the
equipartition value of 2.8 $\times$ 10$^{-4}$ gauss in hot spot A,
calculated assuming no relativistic protons, the broken power-law
spectra of Carilli et al. (1991), low frequency cut-offs at 10 MHz and
high frequency cut-offs at 400 GHz.  The uncertainty in the magnetic
field strength estimated from the SSC model is estimated to be a few
tens of percent.

The most straightforward interpretation is that the relativistic gas
is an electron-positron plasma and is close to equipartition with the
magnetic field. If, on the other hand, the energy in relativistic
protons dominates that in relativistic electrons and the X-rays are
still SSC radiation, the energy density in relativistic protons must
exceed that in the magnetic field. It is notable that the magnetic
field cannot be less than 1.5 $\times$ 10$^{-4}$ gauss since the SSC
radiation would then exceed the observed X-radiation.

The alternative is that B $>$ 1.5 $\times$ 10$^{-4}$ gauss in which
case the predicted SSC emission would be too weak to account for the
observed X-ray emission. The X-rays would then have to be synchrotron
radiation. However, in view of the excellent agreement between the SSC
model and the observations, we consider a synchrotron model
implausible.

\section{Pictor A}
Fig. 4 shows the Chandra image of the nucleus, western jet and western
hot spot of Pictor A (much fainter X-ray emission is found to the east
of the nucleus - see Wilson, Young \& Shopbell 2001). The jet extends
at least 1\farcm9 (110 kpc) westward from the nucleus and is spatially
coincident with the faint radio jet (Perley, R\"oser \& Meisenheimer
1997).  The X-ray jet ``points'' at the western hot spot, which is
4\farcm2 (240 kpc) from the nucleus. When the nuclear X-ray source is
shifted slightly to coincide with the nuclear radio source, the peak
of the X-ray emission from the hot spot is within 1$^{\prime\prime}$
of its radio peak.

Images of the western hot spot at radio, optical (R\"oser 1989;
Perley, R\"oser \& Meisenheimer 1997) and X-ray wavelengths with
similar resolutions are shown in Fig. 5. As may be seen, the overall
X-ray morphology is remarkably similar to the radio and optical. The
high linear polarization of the hot spot at both radio and optical
wavelengths indicates that the emission is synchrotron radiation in
these wavebands.

Fig. 6 shows the broad band spectrum of the western hot spot.  The
radio spectrum of the hot spot is well described by a power law with
$\alpha$ = 0.740 $\pm$ 0.015 (Meisenheimer, Yates \& R\"oser 1997),
but there must be a break or turnover in the spectrum at 10$^{13-14}$
Hz to accommodate the near infrared and optical measurements. It is
also apparent that the X-ray spectrum is not a simple extension of the
radio and optical measurements to higher frequencies.

Wilson, Young \& Shopbell (2001) reached the following conclusions
about the X-ray emission of the hot spot, assuming that its bulk
outward motion is non-relativistic. (i) Inverse Compton scattering of
the synchrotron radio photons by the relativistic electrons
responsible for the radio emission (i.e. a synchrotron self-Compton
model) may be ruled out for the hot spot's X-ray emission, as the
predicted spectrum differs from that observed.  (ii) A successful
inverse Compton model may be constructed for the X-ray emission by
invoking a low energy population of relativistic electrons with an
energy index appropriate to the X-ray spectral index. However, the
magnetic field must be reduced to $\sim$ 1\% of equipartition, and
there is no evidence for such an electron population.  (iii) The
emission may be reproduced in a composite synchrotron plus synchrotron
self-Compton model (solid line in Fig. 6). However, the model is
contrived, requiring similar fluxes from the two components in the
Chandra band. (iv) Synchrotron radiation is possible, but the electron
population must be distinct from that which produces the radio and
optical emission. However, the X-ray spectrum is in excellent
agreement with the expected form if the electrons are accelerated by
strong shocks. Relativistic electrons may also result from a `proton
induced cascade' (e. g. Mannheim, Kr\"ulls \& Biermann 1991).

If the jet is non-relativistic, inverse Compton scattering is an
implausible model for its X-ray emission, for it requires a magnetic
field a factor of 30 below equipartition. Further, it is hard to
understand why the jet is brighter than the lobe (the opposite of the
situation in the radio) in such a model. If the jet is relativistic,
these difficulties are eased, and we consider inverse Compton
scattering by such a jet off the microwave background a viable
mechanism. However, the magnetic field must still be well below
equipartition for plausible angles between the jet and the line of
sight in this lobe-dominated, FRII radio galaxy.  Synchrotron
radiation is another possibility.

\section{The difference between the hot spots of Cygnus A and Pictor A}

It is striking that the X-ray emissions from the hot spots of Cygnus
A, 3C 123 (Hardcastle, Birkinshaw \& Worrall 2000) and 3C 295 (Harris
et al. 2000) conform to an SSC model with a magnetic field close to
equipartition (calculated assuming that only electrons are
present). In three other galaxies - Pictor A, 3C 120 (Harris et
al. 1999) and 3C 390.3 (Harris, Leighly \& Leahy 1998) - the X-ray
emission is orders of magnitude too strong to be SSC emission with an
equipartition magnetic field.  While this is a small sample, we
(Wilson, Young \& Shopbell 2000) noted that 3C 295 and Cygnus A are in
clusters with prominent cooling flows and 3C 123 is in a cluster with
strong, extended X-ray emission and thus may be within a cooling
flow. In contrast, Pictor A, 3C 120 and 3C 390.3 are not in cooling
flows.  Such an environmental difference could affect the X-ray power
from the hot spots in two ways: (a) the presence of high density
surrounding gas may inhibit production of X-ray synchrotron emitting
electrons in hot spots, or (b) the environments of the
``overluminous'' X-ray hot spots might have such low gas density that
the bulk outward motion of the hot spots is relativistic. Luminous
X-ray emission could then originate through inverse Compton scattering
of the microwave background radiation, as we have discussed for the
{\it jet} of Pictor A.  In the second case, objects with high X-ray to
radio flux ratios would tend to be viewed more pole-on than objects
with low X-ray to radio flux ratios.  Chandra observations of a larger
sample of hot spots should cast light on these issues.

This research was supported by NASA grant NAG 81027 and by the
Graduate School of the University of Maryland. We are grateful to Rick
Perley for providing his radio images.


\clearpage

\noindent {\bf Figure Captions}

\bigskip
\noindent Figure 1. A grey scale representation of the Chandra X-ray
image of Cygnus A. The shading is proportional to the square root of
the intensity. Coordinates are for epoch J2000.0 throughout this
paper.

\bigskip
\noindent Figure 2. Chandra X-ray spectrum of the nucleus of Cygnus
A. The x axis represents observed photon energy. The upper panel shows
the data [points with error bars] and a model of an absorbed power law
spectrum folded through the instrumental response (solid line). The
lower panel shows the deviations of the observed spectrum from the
model. Our study of the nuclear spectrum is in collaboration with
Keith Arnaud.

\bigskip
\noindent Figure 3. {\it a:} X-ray emission (contours) superposed on a
6 cm VLA radio map (grey scale) of the region of the western hot spots
(A, the brighter, and B, $\simeq$ 6$^{\prime\prime}$ SE of A) in
Cygnus A. Contours are plotted at 2, 4, 8, 12, 16, 24 and 32 counts
per pixel (0\farcs5 $\times$ 0\farcs5). The grey scale is proportional
to the square root of the radio brightness. {\it b:} Spectrum of hot
spot A.  The points show the radio fluxes and the line through them
the model of the synchrotron radiation. The ``bow tie'' is the Chandra
measured boundary of the X-ray spectrum (these error lines are 90\%
confidence after freezing N$_{\rm H}$ at its best fit value, which
coincides with the Galactic column). The solid line is the predicted
SSC spectrum for $\gamma_{\rm min}$ = 1 and the dashed line for
$\gamma_{\rm min}$ = 100.

\bigskip
\noindent Figure 4. Grey scale representation of the full resolution
Chandra image of the nucleus, jet and western hot spot of Pictor A.

\bigskip
\noindent Figure 5. The morphology of the western hotspot of Pictor~A,
at 3.6 cm radio (left panels; PRM), R band optical (center panels;
R\"oser (1989) and PRM) and X-ray (right panels; this paper)
wavelengths. The upper panels are grey scale representations and the
lower panels are contour plots in which the contours are
logarithmically spaced and separated by a factor of 2. The radio and
optical images have a FWHM resolution $\simeq$ 1\farcs5 (shown as the
filled circles at the bottom right) and the X-ray image has a FWHM of
1\farcs2 $\times$ 1\farcs9 (the PSF is shown at the bottom right; no
smoothing has been applied to the X-ray image).

\bigskip
\noindent Figure 6. The broad band spectrum, plotted as $\nu S_{\nu}$,
of the western hotspot of Pictor~A from radio through X-ray
wavelengths. Note that the scale along the y axis is expanded compared
to that along the x axis. The circular and triangular symbols are
measured data points, while the ``bow tie'' represents the Chandra
spectrum.  The lines represent various models discussed in Wilson,
Young \& Shopbell (2001).

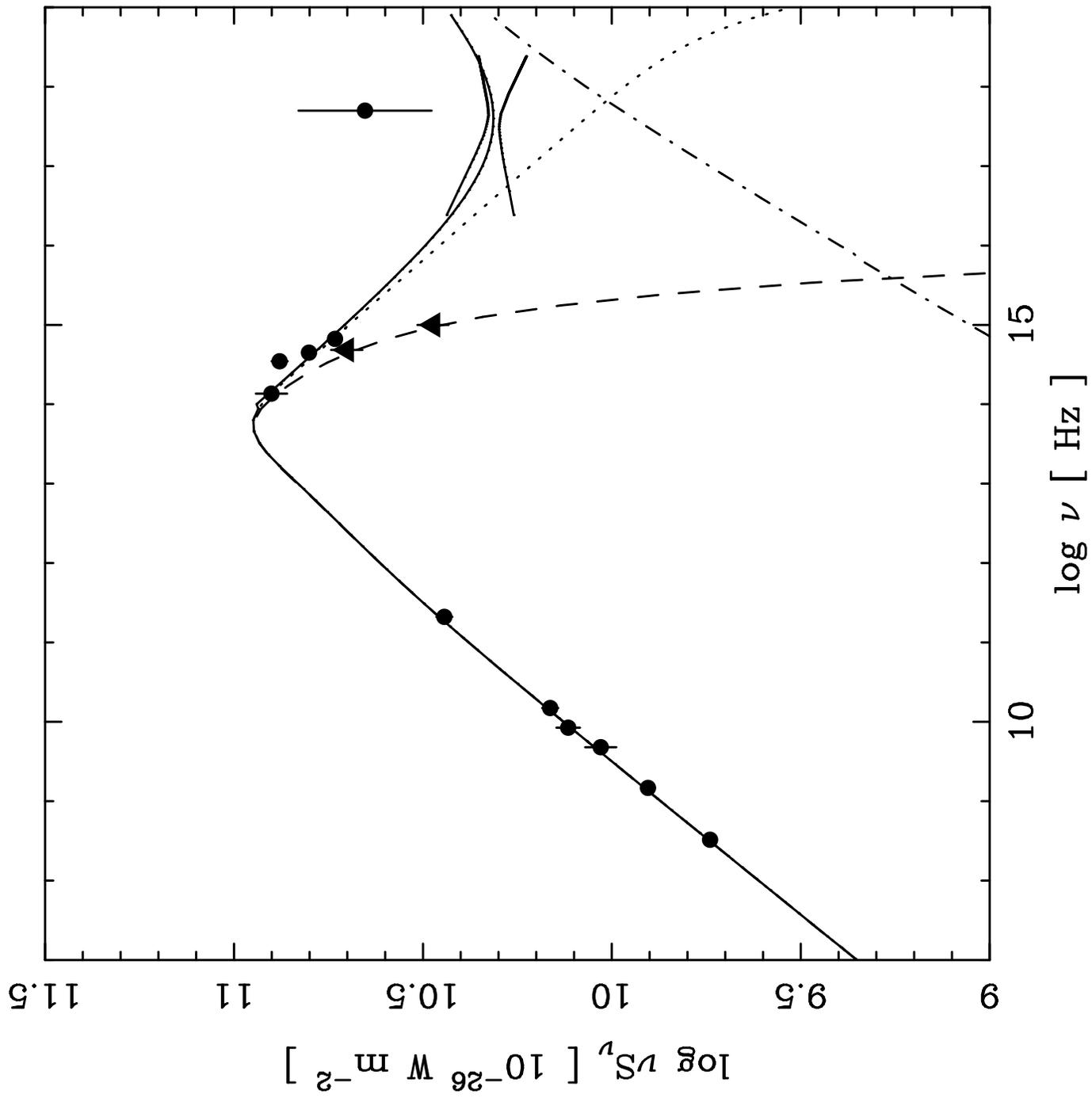
\end{document}